\documentclass[prb,twocolumn]{revtex4}

\usepackage{graphicx}
\usepackage{dcolumn}
\usepackage{bm}

\begin{document}

 \preprint{Am.J.Phys./Ward,Nelson}

\title{Finite Difference Time Domain (FDTD) Simulations of Electromagnetic Wave Propagation Using a Spreadsheet}

\author{David W. Ward}
\author{Keith A. Nelson}
\email{kanelson@mit.edu} \homepage{http://nelson.mit.edu}
\affiliation{%
Department of Chemistry\\Massachusetts Institute of Technology,
Cambridge, Massachusetts 02139}%

\date{\today}

\begin{abstract}
We describe a simple and intuitive implementation of the method of
finite difference time domain simulations for propagating
electromagnetic waves using the simplest possible tools available
in Microsoft Excel. The method overcomes the usual obstacles of
familiarity with programming languages as it relies on little more
than the cut and paste features that are standard in Excel.
Avenues of exploration by students are proposed and sample graphs
are included. The pedagogical effectiveness of the implementation
was tested during an Independent Activities Period class, composed
of 80\verb+%+ freshmen, at MIT, and yielded positive results.
\end{abstract}

\maketitle

\newpage
\section{\label{sec:Intro}Introduction\protect\\}

Here we outline a simple demonstrative example of the method of
finite difference time domain (FDTD) simulations of
electromagnetic wave propagation appropriate for undergraduates in
an introductory electricity and magnetism course or advanced high
school science students using Microsoft Excel, a robust software
package that offers advanced graphics, numeric, and animation
capabilities requiring minimal computer experience.\cite{1,2,3}
Only two rows in the spreadsheet, one for initial electric field
and another for initial magnetic field, need be specified in order
to initiate the simulation, and graphs of the spatially and
temporally evolving waveforms can be updated in real time as the
simulation is carried out in Excel. We begin with a review of the
basic tenets of the FDTD method, proceed into an outline for the
implementation in Microsoft Excel, and conclude with some
suggested exercises.

\section{\label{sec:Methods}Methods}

\subsection{\label{sec:Methods1}FDTD basics}

The starting point for an FDTD simulation are Maxwell's equations,
which are repeated here for the case of one dimensional free space
propagation in time (\emph{t}) and space (\emph{z}) with no
sources or sinks for magnetic or electric fields \textbf{B} or
\textbf{E} respectively for the corresponding material responses
\textbf{H} or \textbf{D},
\begin{equation}
{
 \frac{\partial E_x}{\partial t}=-\frac{1}{\epsilon_0}\frac{\partial H_y}{\partial
 z},
} \label{maxwell1}
\end{equation}
\begin{equation}
{
 \frac{\partial H_y}{\partial t}=-\frac{1}{\mu_0}\frac{\partial E_x}{\partial
 z},
} \label{maxwell2}
\end{equation}
\begin{equation}
{ \nabla\cdot E=0, } \label{maxwell3}
\end{equation}
\begin{equation}
{ \nabla\cdot B=0. } \label{maxwell4}
\end{equation}

If we choose the x-direction for the polarization of the electric
field and the z-direction for direction of propagation, then it
follows that the magnetic field is y-polarized as indicated in the
subscripts on the fields above. To discretize the equations of
propagation, central difference approximations are taken for the
derivatives of time and space. Temporal steps are indexed by the
integer \emph{n} and related to continuous time by the relation
$t=n\triangle t$, and spatial steps are indexed by the integer
\emph{k} and related to continuous space by the relation
$z=k\triangle z$. The temporal discretization method using central
differences can then be designated for variable \textbf{X} as

\begin{equation}
{
 \frac{\partial X}{\partial t}\longrightarrow \frac{\Delta
 X}{\Delta
 t}\equiv\frac{X(n+1/2)-X(n-1/2)}{\Delta t},
} \label{distime}
\end{equation}

\noindent and spatial discretization as

\begin{equation}
{
 \frac{\partial X}{\partial z}\longrightarrow \frac{\Delta
 X}{\Delta
 z}\equiv\frac{X(k+1/2)-X(k-1/2)}{\Delta z}.
} \label{disspace}
\end{equation}

In FDTD, we need only consider the two curl equations
\ref{maxwell1} and \ref{maxwell2} above because the divergence
conditions \ref{maxwell3} and \ref{maxwell4} can be satisfied
implicitly by interleaving the electric and magnetic field
components in space in what has come to be known as a Yee
cell.\cite{4} A consequence of the spatial interleaving is that
the fields must also be interleaved in time, known as "leapfrog",
since the temporal response of one field is proportional to the
spatial variation of the other at the previous time step.
Implementing eqs. (\ref{maxwell1}) and (\ref{maxwell2}) as
indicated in eqs. (\ref{distime}) and (\ref{disspace}) and
interleaving spatially and temporally yields the difference
equation version of Maxwell's equations:

\begin{equation}
{ \frac{E_x^{n+\frac{1}{2}}(k)-E_x^{n-\frac{1}{2}}(k)}{\Delta
t}=-\frac{H_y^n(k+\frac{1}{2})-H_y^n(k-\frac{1}{2})}{\epsilon_0\Delta
z}
 },
 \label{dismaxwell1}
\end{equation}
\begin{equation}
{ \frac{H_y^{n+1}(k+\frac{1}{2})-H_y^n(k+\frac{1}{2})}{\Delta
t}=-\frac{E_x^{n+\frac{1}{2}}(k+1)-E_x^{n+\frac{1}{2}}(k)}{\mu_0\Delta
z}
 }.
 \label{dismaxwell2}
\end{equation}

\noindent At a given position in space, the field at each current
time step can be calculated from the previous values of the
fields. Solving for the latest field at the latest time step in
eqs. (\ref{dismaxwell1}) and (\ref{dismaxwell2}) yields:

\begin{eqnarray}
E_x^{n+\frac{1}{2}}(k)=\nonumber\\E_x^{n-\frac{1}{2}}(k)-\frac{\Delta
t}{\epsilon_0\Delta z}
\{H_y^n(k+\frac{1}{2})-H_y^n(k-\frac{1}{2})\} , \label{Enew}
\end{eqnarray}

\begin{eqnarray}
H_y^{n+1}(k+\frac{1}{2})=\nonumber\\H_y^n(k+\frac{1}{2})-\frac{\Delta
t}{\mu_0 \Delta z}\{E_x^{n+\frac{1}{2}}(k+1)
-E_x^{n+\frac{1}{2}}(k) \} . \label{Hnew}
\end{eqnarray}

In this fashion, known as Euler forward, the solution of Maxwell's
equations proceeds much the same way that we envision
electromagnetic wave propagation{--}an electric field induces a
magnetic field, which induces and electric field, ad infinitum.
The time step and grid size parameters are chosen based on the
propagating wave frequency and wavelength. For stability, the
general rule is that at least ten grid points sum to less than the
smallest wavelength $\lambda_{min}$ considered, and the Courant
condition then determines the time step:\cite{5}

\begin{equation}
{ \Delta z\leq\frac{\lambda_{min}}{10}
 },
 \label{lammin}
\end{equation}

\begin{equation}
{ \Delta t\leq\frac{\Delta z}{2c_0}
 }.
 \label{courant}
\end{equation}

\noindent where $c_0$ is the speed of light in vacuum.

The computer algorithm implementation of this procedure, eqs.
(\ref{Enew}) and (\ref{Hnew}), requires two one-dimensional arrays
in the spatial coordinate, one for $E_x$ and one for $H_y$, to be
allocated in memory. After imposing an initial condition, in the
form of an initial electric or magnetic field, each time step is
calculated by the following prescription:

\begin{eqnarray}
t=t_0+n\frac{\Delta t}{2}\nonumber\\
\tilde{E}_x[k]=\tilde{E}_x[k]+\frac{1}{2}\{H_y[k-1]-H_y[k]\}\nonumber\\
t=t_0+n\frac{\Delta t}{2}\nonumber\\
H_y[k]=H_y[k]+\frac{1}{2}\{\tilde{E}_x[k]-\tilde{E}_x[k+1]\}
\label{algorithm}
\end{eqnarray}

\noindent where we have adopted the normalized fields
($\tilde{E}=\sqrt{\epsilon_0\mu_0}E$) to simplify the code, and
included the stability conditions, eqs. (\ref{lammin}) and
(\ref{courant}), within the constant preceding the curl.\cite{6}
The algorithm (eqn. (\ref{algorithm})) is iterated for the desired
number of time steps. Note that iteration of (\ref{algorithm})
results in an implicit time formulation, i.e. time is not made
explicit in the equations for the field and only appears in the
algorithm above for bookkeeping purposes.

\subsection{\label{sec:Methods2}FDTD spreadsheet algorithm}
The FDTD methodology for one dimension can be implemented in
spreadsheet format, here in Microsoft Excel, using simple cell
formulas and cut-and-paste features. This has an advantage over
other pedagogical approaches\cite{7} in that no programming
experience is required. The starting point is the algorithm in
(\ref{algorithm}) where columns represent spatial steps and pairs
of rows (one for E and one for H) represent time steps. The
algorithm is illustrated graphically in the center section of
figure 1. For any time step (any pair of rows of $E$ and $H$), the
mapping of (\ref{algorithm}) is: $E_x(column)/H_y(column)$. Since
the electric field value from one prior cell is needed to compute
the curl, $E_x$ begins at column one and ends at column $k+1$.
Similarly, the magnetic field begins at column zero but ends at
column $k$, since its value at one subsequent cell is needed when
computing the curl. The first two rows ($t=0$) are for the initial
conditions. The algorithm computes the remaining cells. The
student need only type in the formulas for $E$ and $H$ for the
second time step and the first two columns; cut and paste may be
used to complete the remaining spatial columns and temporal rows.

The procedure for implementing the algorithm is as follows:
\begin{enumerate}
    \item As a visual aid, enter the spatial coordinate (1, 2, 3, etc) in the topmost row of the spreadsheet starting at column $B$. After typing in the first two spatial coordinates, select cells $B$ and $C$ and drag the "fill" handle to encompass as large a problem space as needed.
    \item Type 'Ex0' in cell $A2$ and 'Hy0' in cell $A3$, where the number following the field component refers to the time step. Drag the "fill" handle down the column, as in 1, to include the desired number of time steps.
    \item Highlight the leftmost and rightmost columns of the problem space.
    \item Fill in all of row 2 and 3 (time step 0) to the end of the problem space with zeros and highlight to indicate that this is the initial condition for the fields.
    \item In the leftmost column and first time step for Ex, type '=B2'. Do the same for the rightmost column. This imposes a perfect electric conductor radiation boundary condition on the problem space, which means that impinging waves will reflect from the problem space boundaries as they would from a mirror.
    \item In the second spatial position of Ex for time step one (cell $C4$), type the formula '=C2+0.5*(B3-C3)' and press enter.
    \item Select the cell from the previous step and drag the "fill" handle to the last un-highlighted spatial column in the row.
    \item In the first spatial position of $H_y$ for time step one (cell $C5$) type the formula '=B2+0.5*(B4-C4)' and press enter. These last two steps define the computing steps of the algorithm.
    \item Select the cell from the previous step, and drag the "fill" handle to the last un-highlighted spatial column in the row.
    \item Finally, select all the columns for time step 1 ($E_x$ and $H_y$), and drag the "fill" handle through the last time step in the simulation. Test that it works by entering a 1 in the initial condition regions and see that all cells automatically update; note that the only valid regions for initialization are at the zero time step ($E_x$ and/or $H_y$) and between the first and last columns of the problem space (highlighted).
\end{enumerate}

The FDTD Excel code will now update all fields in response to the
initial conditions entered by the user. To graphically visualize
the simulation output, a single row can be selected and graphed
using the \textbf{insert $>$ chart $>$ (xy)scatter} menu item.
This outputs the spatial field pattern at the moment in time
designated by the row number. To graph temporal evolution at a
single point in space, it is recommended to make a new field mesh
with only one field component (either $E_x$ or $H_y$) on a
separate sheet and graph from that; otherwise, both the E and H
fields will appear in the graphs. This can be done using
cut-and-paste. An example spatial and temporal graph of two
counter-propagating pulses is illustrated in figure 3.

\section{\label{sec:results}Applications of the algorithm}

\subsection{\label{sec:Apps1}Power flow}

The direction of propagation of light is dictated by the phase
between the electric and magnetic fields. For transverse
electromagnetic (TEM) wave propagation the familiar right-hand
rule applied by curling $E$ into $H$ indicates the direction of
power flow. Using our spreadsheet code, this can be demonstrated
by initializing both the electric and magnetic field as a
sinusoidal waveform modulated by a Gaussian envelope. Introducing
a phase shift of either $0$ or $\pi$ in the magnetic field allows
the direction of propagation to be controlled. If the electric
($E_x$) and magnetic ($H_y$) field are both positive at the same
points in space and time, then the pulse travels in the positive
$z$ direction; if the signs are opposite then propagation is in
the opposite direction as illustrated in figure 2.

\subsection{\label{sec:Apps2}Guided wave optics}

Some of the most interesting applications of electrodynamics occur
when constraints are placed on the fields. The perfect electric
conductor radiation boundary conditions form the walls of a
resonator when the wavelength in the problem space (resonator
width) is of similar size to the problem space itself. With PEC
walls, no light wave with a wavelength longer than twice the
resonator width will propagate. This is referred to as a cut-off
wavelength. Also, since the electric field must be zero at the
boundary and only nodes in a plane wave may satisfy this, only
wavelengths that are half-integer multiples of the resonator width
can propagate. Resonator modes can be investigated by setting the
initial conditions to a plane wave with nodes at the boundaries is
illustrated in figure 3.

\subsection{\label{sec:Apps3}Transmission and reflection}

By introducing a relative permittivity $\varepsilon_r$ into
equation (\ref{Enew}), which changes the index of refraction and
hence the wave propagation speed, the spreadsheet code can
illustrate reflection and transmission at a material interface. In
terms of the normalized fields, the constant $1/2$ becomes
$1/2\varepsilon_r$. Small errors in the transmitted and reflected
fields are to be expected, but these will decrease as the number
of mesh points per wavelength increases. Figure 4 illustrates this
process for an air ($\varepsilon_r=1$)/material
($\varepsilon_r=2$) interface.

\section{\label{sec:Activities}Further exercises and suggested activities}

The following exercises are recommended. The spreadsheet code for
the examples is available online, but it is recommended that
students augment the code on their own, because writing and
tinkering with the code is the best way to understand how it
works. Exercises are listed in order of increasing difficulty, or
in a manner such that each exercise depends only on changes
already administered.
\subsection{\label{sec:Activities1}Fundamental}

\begin{enumerate}
    \item Use trigonometric and exponential mathematical functions to generate the initial conditions and reproduce the results in figure 2.
    \item Study the cavity modes of a one-dimensional resonator, figure 3, by introducing standing waves into the initial conditions. What happens when the wavelength becomes longer than the problem space size (the width of the resonator)?
    \item Experiment with radiation boundary conditions. Change the radiation boundary condition from perfect electric conductor to perfect magnetic conductor (H$_y$ is zero at the problem space boundaries). What is different about reflections from these two kinds of surfaces?
    \item Implement a periodic boundary condition for unbounded propagation.
    \item Reproduce the reflection and transmission simulation in figure 4, and then change the interface from air/material to material/air. What changes?
\end{enumerate}

\subsection{\label{sec:Activities2}Advanced}

\begin{enumerate}
    \item Experiment with spatially periodic variation in the relative permittivity. This should produce frequency-dependent filtering, and for some conditions should produce "photonic bandgap" materials with no propagation in certain frequency regions.
    \item Introduce loss into the simulation by including non-zero conductivity in Maxwell's equations and implementing it into the FDTD code.
    \item Implement an absorbing boundary condition.
\end{enumerate}

\section{\label{sec:Conclusion}Conclusion}
We have formulated a simple implementation of the FDTD method of
propagating electromagnetic waves using basic features available
in Microsoft Excel, and we have presented some illustrative
examples. The implementation was tested on freshmen at MIT during
its 2004 Independent Activities Period in a short course\cite{8}
that met for two hours on each of two consecutive days. This brief
introduction was sufficient for students to gain substantial
proficiency in elementary simulations like those illustrated
above. The class enjoyed unusually high attendance, especially
among students preparing for or recently involved in
electromagnetism coursework. Instructor (D.W.W.) observation and
student feedback after the course indicated considerable
satisfaction with insights and proficiency gained. Several student
responses pinpointed the graphical form of the spreadsheet based
algorithm as the key merit of the approach. The benefit of keeping
track of a small number of unit cells in order to see how the
algorithm works is useful for beginning and intermediate students
of computer simulation, numerical recipes, and electricity and
magnetism. The automatic updating of Excel graphs makes this an
effective learning tool for students interested in the basics of
electromagnetic wave propagation, as the consequences of changes
in initial conditions are illustrated graphically and instantly.
Finally, the methodology learned in this rapid introduction may
stimulate student interest in more advanced FDTD simulation
techniques and their broad research applications.\cite{9,10}
\begin{acknowledgments}
This work was supported in part by the National Science Foundation
Grant no. CHE-0212375 and MRSEC award no. DMR-0213282, and by the
Cambridge-MIT Institute Grant no. CMI-001.
\end{acknowledgments}

\bibliography{FDTD_with_Excel}
\end{document}